\documentclass[aps,prl,twocolumn,showpacs,10pt]{revtex4-2}
\usepackage[utf8]{inputenc}
\usepackage{amsmath,amssymb}
\usepackage{amsfonts}
\usepackage{amsthm}
\usepackage{hyperref}
\usepackage{graphicx}
\usepackage{xcolor}
\usepackage{mathtools}
\usepackage{verbatim}
\usepackage{float}
\usepackage{comment}

\begin{document}
\title{Scaling laws and relaxation rate at desynchronization}
\author{Ayushi Suman}
\email{ayushishuman@gmail.com}
\author{Sarika Jalan}
\affiliation{Complex Systems Lab, Department of Physics, Indian Institute of Technology Indore, Khandwa Road, Simrol, Indore-453552, India}

\date{\today}

\begin{abstract}
While divergence in length and time, hallmarks of criticality, has been established for second order phase transition to synchronization in the Kuramoto model, no theoretical investigations exist on such scaling associated with abrupt transitions. Through numerical simulations and analytical derivations at the abrupt desynchronization transition for the Kuramoto model with higher-order interactions, 
we document the emergence of critical slowing down, which is marked by a diverging relaxation time. We verify a finite-time finite-size scaling by applying the homogeneity assumption and defining a correlation in a set endowed with counting measure. 
This work extends the theoretical foundation of critical scaling to include discontinuous phase transitions in coupled oscillator systems.

\end{abstract}

\maketitle
\paragraph{Introduction:}
The Kuramoto model is a framework used to study synchronization in real-world systems. Developed by Y. Kuramoto \cite{kuramoto1987statistical} during the 1970s 
the model retains the key features of collective synchronization in a comparatively simple mathematical form. Despite of its simplicity, the Kuramoto model has been applied extensively in a variety of areas, ranging from neuroscience and circadian biology to power grid dynamics and social systems \cite{strogatz2000kuramoto}.
In its traditional form, the Kuramoto model displays an extensively characterized continuous synchronization transition \cite{daido1990intrinsic}. Above a critical coupling strength, the system undergoes a transition from an incoherent state where oscillators rotate with their natural frequencies to a partially synchronized state where a proportion of oscillators lock to a common frequency. This transition takes the normal form of a supercritical pitchfork bifurcation.
Recent studies have emphasized an extension of the traditional Kuramoto model to incorporate higher-order interactions. This is stimulated by the understanding that many real-world complex systems consist of interactions that cannot be properly described within the framework of pairwise coupling only. For example, in biological neural networks, synaptic couplings interact with more than one neurons at a time \cite{ravid2016shaping}. In social dynamics, opinion formation can rely on influence from more than pairwise interactions \cite{battiston2020networks}. Likewise, in technological networks, the action of an individual node can be driven by the common state of several neighboring nodes \cite{joshi2012autonomous}. An inclusion of higher-order interactions in Kuramoto model \cite{skardal2020higher} have shown to change the supercritical bifurcation to a combination of saddle node and subcritical bifurcations, thereby creating an irreversibility and hysteresis with quasi-static change in the pairwise coupling acting as a control parameter. Consequently, a continuous transition to synchronization changes into two abrupt transitions creating a hysteresis. 
While the second order phase transition in Kuramoto model displays universal scaling behavior \cite{hong2015finite, xu2020universal, dorfler2011critical}, and algebraic dependencies have been characterized for first order transitions \cite{wang2021phase,pazo2005thermodynamic}, critical phenomena with divergent time scales at abrupt transitions remain unexplored.
This Letter shows that the the abrupt transition from synchronous to asynchronous state with an adiabatic decrease in the  coupling yields the sign of critical slowing down marked by diverging correlation length and time scale. 
Divergent time scales at a first order transition has been studied in spatially interacting systems and has been termed mixed-order transition\cite{korbel2025microscopic,bar2014mixed}. On contrary, here we observe such divergences due to interactions in a number space which does not hold any concept of locality.

The mathematical formalism of scaling analysis has turned out to be a valuable
approach to the description of collective behavior in many-body systems in a wide range of
fields of physics and applied mathematics \cite{stanley1987introduction} \cite{kadanoff2000statistical}. Power-law relations, finite-size scaling methods,
and recognition of universal scaling functions allow one to determine characteristic exponents
describing quantitative system behavior in the vicinity of transition points. Scaling analysis
methodologies are well developed in equilibrium statistical mechanics and critical phenomena \cite{kadanoff2000statistical}.
Their systematic extension to non equilibrium dynamical systems like coupled oscillators provides valuable insights into understanding collective dynamics and relaxation processes in complex systems \cite{hong2015finite,kuramoto1995scaling,lee2014finite, xu2020universal}.  
Finite-size scaling theory generalizes these ideas to finite systems, allowing extrapolation to thermodynamic limits via data collapse methods. Here we establish scaling relations and demonstrate divergence of the relaxation time at the desynchronization threshold in Kuramoto models with triadic interactions. We also confirm a finite-time finite-size scaling (FTFSS) with the homogeneity assumption \cite{stanley1987introduction} and taking into account correlations not defined in a Lebesgue measure set but rather in a set endowed with counting measure.

\paragraph{Model:}
We consider the Kuramoto model \cite{kuramoto1987statistical} consisting of a $N$ number of phase oscillators interacting via a non-linear pair-wise coupling with strength $k$ acting as a control parameter. Strength of three body (triadic) interactions among oscillators is denoted by $k_h$. The oscillators themselves have an intrinsic frequency which is sampled from a Lorentzian distribution $\omega_{i \in I}$ where $I = \{1, 2, 3, \dots, N\}$ and have respective phases $\theta_{i\in I}$. The coupled dynamics under this description is given by 
\begin{figure*}
\includegraphics[width=\textwidth]{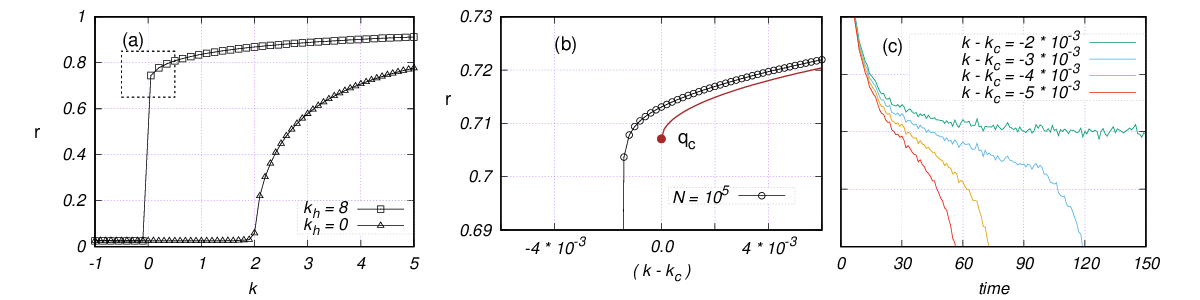}
\caption{\textbf{Numerical observations:} (a) Dependence of $r_1$ on $k$ for various $k_h$ values and $N = 10^5$ is shown; a second-order transition changes into first order due to inclusion of higher-order interactions. (b) The effect of critical slowing down is depicted for $N = 10^5$ which shows deviation from the analytical curve (brown solid line) obtained by solving (\eqref{r1(q)}) and (\eqref{r2(q)}) simultaneously and $q_c$ is the analytical transition point. (c) Critical slowing down for negative $(k - k_c)$ is shown close to $(k-k_c)=0$. As we advance towards the transition point, the relaxation time $\tau$ to reach the equilibrium state increases.  }
\label{figure1} 
\end{figure*}
\begin{equation}
    \theta_i = \omega_i + \frac{k}{N} \sum_{j \in I} sin(\theta_j - \theta_i) + \frac{k_h}{N^2} \sum_{j,k \in I} sin(2\theta_j - \theta_k - \theta_i).
\label{model}
\end{equation}
To quantify the extent of synchronization, Kuramoto proposed the use of a complex-valued quantity which is equal to the centroid of the instantaneous phases of all the oscillators in the complex plane as
\begin{equation}
 z_m= r_me^{\iota \psi_m} = \frac{1}{N}\sum\limits_{j=1}^{N} e^{\iota m\theta_i}.
\label{eq_order_parameter}
\end{equation}
The subscript for $m=1$  will hereafter be omitted for notational ease, ($z_1 \to z$) and ($r_1 \to r$). Therefore Eq ~\eqref{model}, using Eq \eqref{eq_order_parameter}, reduces to mean-field  which in steady state $(\dot{\theta_i} \to 0)$ yields conditions for phase locking after taking an average over all the oscillators
\begin{equation}
    \langle \omega_i \rangle_I = k r \langle sin(\theta_i-\psi_i) \rangle_I + k_h r_2 r \langle sin( \theta_i - (\psi_2 - \psi_1))\rangle_I.
    \label{mean_field}
\end{equation}
What follows that \eqref{model} carries a rotational symmetry $\theta_i \mapsto \theta_i + c$ and therefore without loss of generality, $g(\omega)$ can be taken to be centered at zero. This Letter considers $g(w)$ to be Lorentzian with mean $\omega_0 = 0$ and standard deviation $\Delta = 1$, 
for which \eqref{mean_field} provides the condition for phase locking as 
\begin{equation}
    \langle sin(\theta_i-\psi_i) \rangle_I = -\frac{k_h}{k} r_2 \langle sin( \theta_i - (\psi_2 - \psi_1))\rangle_I. 
    \label{phase_locking_condn}
\end{equation}
Next, $\forall \quad i \in I, \: \exists \quad j \in I$ such that $sin\:\theta_i = - sin\: \theta_j$, where $\theta_i, \theta_j \: \in [0,2\pi)$ is a solution to the condition in \eqref{phase_locking_condn}, which implies $\psi_1 = \psi_2 = 0$, and the oscillators density in the steady state follows
\begin{equation}
    \rho(\theta \: ;\omega, t \to \infty) = 
    \begin{cases}
        \delta(\theta - sin^{-1}\frac{\omega}{q} )\Theta(cos \: \theta) \quad \quad |\omega| \leq q \\
        \frac{C}{|\omega - q\:sin\:\theta|} \quad\quad\quad\quad\quad\quad\quad\quad |\omega| > q, 
    \end{cases}
\label{density_oscillators}
\end{equation}
where $q = k r  + k_h r_2 r$, $\Theta(cos\:\theta)$ is the heavy-side step function and the normalization constant $C = \frac{1}{2\pi}\sqrt{\omega^2 - q^2}$. Hence, depending on the natural frequencies, the oscillators can be divided into two groups, the ones that are locked to the mean-field with $sin\:\theta_i = \frac{\omega_i}{q}\:(|\omega_i| \leq q)$ and the ones that drift uniformly with $|\omega_i| > q$. The order parameters are then given by $z = r = \langle e^{\iota\theta} \rangle_{lock} + \langle e^{\iota\theta} \rangle_{drift}$ and $z_2 = r_2 = \langle e^{\iota 2\theta} \rangle_{lock} + \langle e^{\iota 2\theta} \rangle_{drift}$.

Under the symmetry $g(\omega) = g(-\omega)$ and $\rho(\theta+\pi \:; -\omega) = \rho(\theta\: ;\omega)$ implied by Eq~\eqref{density_oscillators}, $\langle e^{\iota\theta} \rangle_{drift} = 0$, and the magnitude of order parameters  in Eq~\eqref{eq_order_parameter} in the $N \to \infty$ limit can be written in an integral form
\begin{equation}
    r = \int_{-q}^q \sqrt{1-\left(\frac{\omega}{q}\right)^2} g(\omega) d\omega \quad \equiv \:R(q),
\label{r1(q)}
\end{equation}
\begin{align}
    r_2 =  \int_{-\pi}^{\pi}\int_{-\infty}^\infty e^{2\iota\theta} \rho(\theta\: ;\omega)g(\omega)\:d\theta\:d\omega \quad 
    \equiv \: R_2(q),
\label{r2(q)}
\end{align}
where $R_2(q)$ contains both locked and drifting contributions. Therefore the macroscopic quantities $r$ and $r_2$ can be obtained as a function of $q$ alone.

\paragraph{Simulating the model:}
The model ~\eqref{model} is evolved numerically  with 'RK4' method for uniformly distributed Lorentzian frequencies. The initial condition for these simulations is taken to be $\theta_i(0) = 0\; \forall i \in I$ and the system of ODEs is then evolved by adiabatically decreasing $k$. As $k_h$ increases, the second-order phase transition changes into a first-order abrupt transition. Fig.~\ref{figure1}(a) shows that starting from the random phases, for $k_h=0$,  that is, in the absence of triadic interactions, the coupled oscillators display usual second-order phase transition. Upon introducing triadic interactions ($k_h>k_c)$, with an adiabatic decrease in $k$, an abrupt transition from synchronous to desynchronous state takes place \cite{skardal2020higher}. Fig.~\ref{figure1}(b) presents a zoomed-in view of ~\ref{figure1}(a) near desynchronization point which shows that the simulated results do not align with the analytical curve obtained by solving Eq~\eqref{r1(q)} and Eq~\eqref{r2(q)} in the vicinity of $q_c$.  This deviation is due to the phenomenon of {\textit critical slowing down} in which the relaxation time of the system to reach the steady state diverges near a critical point (Fig.~\ref{figure1}(c)).
The relaxation time to reach equilibrium state keeps on increasing  as we advance towards the critical point (Fig.~\ref{figure1}(c)) from the negative coupling side. For $\delta k = -5 * 10^{-3} $, the system relaxes to equilibrium in $\sim 40$ seconds while for $\delta k = -2 * 10^{-3} $, which is much closer to $k_c$, the system does not relax to equilibrium even in over $\sim 150$ seconds.

\paragraph{Critical scaling exponents:}
Characterization of the transition point begins with the identification of scaling exponents and the relationships among them. In calculations of bulk scaling exponents  at $q_c$, both leading and the sub-leading terms  will be considered to account for the effects of their crossover at finite distances from the critical point. To this end, we define a characteristic function $F(q)$ using $q = k r + k_h r_2r$ as
\begin{equation}
    k = \frac{q - k_h \:R(q)\:R_2(q)}{R(q)} \equiv F(q).
\label{define_characteristic function}
\nonumber
\end{equation}
At bifurcation points, $\frac{\partial r}{\partial k}$ diverges, hence $\frac{\partial r}{\partial q} (\frac{\partial k}{\partial q})^{-1}$ diverges and $\frac{\partial F(q)}{\partial q} \; \to 0$. $F(q)$ has local extrema at all bifurcation points.

\begin{figure}
    \centering
        \includegraphics[width=0.49\textwidth]{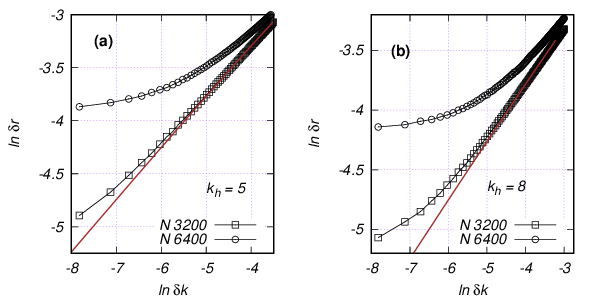}
        \includegraphics[width=0.49\textwidth]{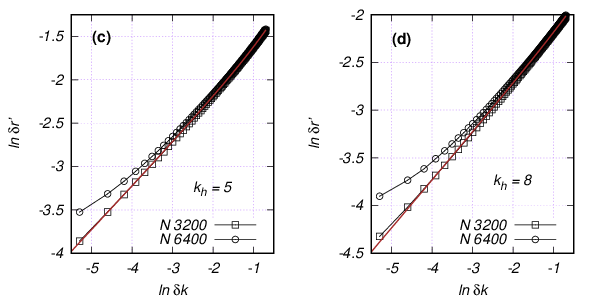}
        \caption{\textbf{Deviation from bulk scaling:} (a) and (b) disclose deviation of numerical observation from the analytical bulk scaling (brown solid line, \eqref{scaling_r1}) with leading term alone for two different system size for $k_h = 5$ and $k_h =8$, respectively. (c) and (d) show deviation when subleading contribution is also taken into consideration using ~\eqref{scaling_r1}. }
    \label{figure2} 
\end{figure}

Taking a perturbation $\delta q \ll 1$ in the characteristic parameter at $q_c$ and expanding $F(q_c + \delta q) = F(q_c) + \delta[F(q)]$, we get
\begin{equation}
    \delta k = \delta[F(q)] = X\: \delta q^a + Y\: \delta q^b; \quad a < b,
\label{delta_variation}
\end{equation}
where $a$ and $b$ are the lowest surviving exponents in the Taylor expansion of $F(q)$.
\eqref{delta_variation} can be inverted as
\begin{equation}
    \delta q = X^{-1/a} \: \delta k^{1/a} + H\: \delta k ^c
    \nonumber
\end{equation}
with $H = -\frac{Y}{a} X^{-(b + 1)/a}$, $c= \frac{b}{a} + \frac{1}{a} -1$ and $\delta k = k - k_c$ is the reduced coupling.
For a Lorentzian frequency distribution, Eq \eqref{r1(q)} and \eqref{r2(q)} can be integrated and expansion of $F(q)$ gives $a = 2$, $b = 3$ and $c = 1$. Next, using $q = k\:r + k_h\:r_2\:r$, a variation in $r$ can be obtained as (Appendix 1)
\begin{equation}
    \delta r = C_1 \:\: \delta k^{1/2} + C_2 \: \: \delta k, \:
\label{scaling_r1}
\end{equation}
where $C1(r_c, k_c, r_{2,c};k_h)$ and $C2(r_c, k_c, r_{2,c};k_h)$ are leading and subleading coefficients. Scaling exponents for $r_2$ can be realized similarly \cite{supplementary}. 
Figs.~\ref{figure2}(a) and \ref{figure2}(b) depict the behavior of the leading scaling exponent and deviations of numerical data from analytical predictions for various combination of $N$ and $k_h$. 
For scaling up to the sub-leading order, we analyze $\delta r' = C_1 \: \delta k^{1/2}$, ($\delta r' = \delta r /\{1 + (C_2 / C_1) \delta k ^{1/2}\})$ on a log-log graph (Figs.~\ref{figure2}(c) and ~\ref{figure2}(d), for two different values of $k_h$) which follow a straight line for subleading order scaling. Deviations from the analytical prediction  \eqref{scaling_r1} are observed. As expected, the deviation due to critical slowing down is pronounced in the leading as well as subleading terms. The deviation is larger for larger $N$ which is a clear indicator of divergence in the time scale. This will be quantified in subsequent sections.

\begin{figure*}
\centering
\includegraphics[width=\textwidth]{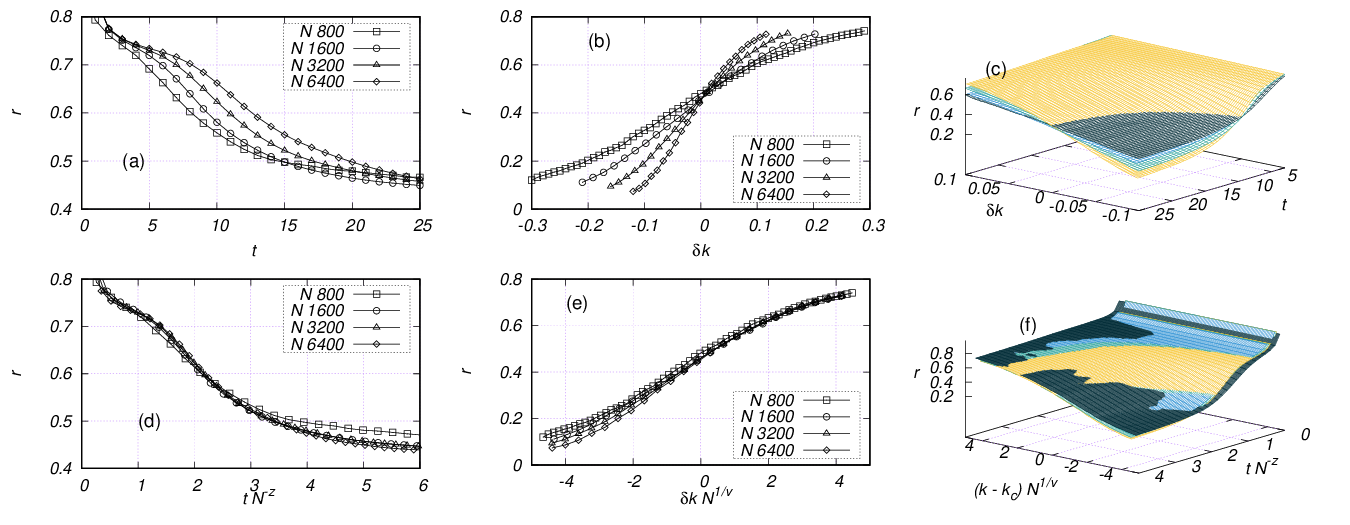}
\caption{\textbf{FTFSS:} $r$ for different system sizes as a function of (a) time by keeping $\delta k = 0$, (b) reduced coupling $\delta k$ by keeping $N^{-\eta / \nu} t = 5$ fixed (b). (c) 3D surface plot showing dependence of $r$ on time and reduced coupling. (d) and (e) are scaled graphs showing data collapse when plotted against scaled time $t N^{-\eta / \nu}$ (at 
fixed $\delta k = 0$) and scaled coupling $\delta k N^{1 / \nu}$ (at fixed $N^{-\eta / \nu} t = 5$), respectively. (f) 3D surface plot in scaled variables. Note that this figure utilizes a random sampling from $g(\omega)$, a variation from the uniform sampling employed in preceding figures. Consequently, observed deviations in $r_c$ and $k_c$ is attributed to this sampling method.}  
\label{figure3} 
\end{figure*}

\paragraph{Derivation of relaxation rate:}
A theorem in harmonic analysis states that if the density function $\rho( \theta \: ; \omega,t)$ has analytic continuation in an open disk $ \mathcal{D} := \{ z \in \mathbb{C} : \: 1-\epsilon  < |z| < 1+\epsilon\}$,  each frequency mode of the Fourier expansion of $\rho( \theta \: ; \omega,t)$ decays exponentially \cite{katznelson2004introduction}, 
  $  \hat{\rho_n} = \alpha^n(\omega,t) $
with $|\alpha| < 1$ for drifting oscillators and $|\alpha| \to 1$ for locked oscillators.  The distribution function $\rho( \theta \: ; \omega,t)$ for a given $\omega$ takes the form of the Poisson kernel multiplied by $g(w)/2\pi$. Using this distribution in the continuity equation and averaging over $g(\omega)$ leads to reduction of the infinite dimensional dynamics to a one dimensional ODE in the macroscopic observable $r$ for a Lorentzian distribution of natural frequencies \cite{supplementary},
\begin{equation}
    \dot{r} = f(r; k)
    \label{rdot}
\end{equation}
where \( f(r_s; k) = 0 \) with \( r_s \) representing a stable fixed point of Eq~\eqref{rdot}, indicates an equilibrium macrostate of the system in the thermodynamic limit.
Applying a linear perturbation $ \delta r(t) = r(t) - r_s$ in the equilibrium state, the dynamics near $r_s$ can be linearized as (Appendix 2),
\begin{equation}
    \frac{d}{dt} \delta r = f'(r_s;k)\: \delta r,
    \nonumber
\end{equation}
giving an exponential decay of $\delta r$ as
$\delta r(t) = \delta r(0) e^{-t/\tau}$,
where $\tau = [f'(r_s;k) ] ^{-1}$ is the relaxation time. This quantity can be shown to diverge by expanding $f'(r_s;k)$ near $k_c$,
\begin{align*}
    f'(r_s;k) &= f'(r_c;k_c) +\left. \frac{\partial f'}{\partial r} \right|_{r_c, k_c} (r_s - r_c)\\ \nonumber
    & \quad \quad \quad \quad \quad \:\:+ \left. \frac{\partial f'}{\partial k} \right|_{r_c, k_c} (k - k_c) + \mathcal{O}(\delta r ^2, \delta k^2)\\ \nonumber
    &= 0 + f''(r_c;k_c)\delta r + \left. \frac{\partial f'}{\partial k} \right| \delta k \: \nonumber.
\end{align*}
As $\delta k \to 0$, the term $f''(r_c;k_c)\delta r$ scales as $\sqrt{\delta k}$ (Eq~\eqref{scaling_r1}) and dominates the term proportional to $\delta k$. Therefore,
\begin{equation}
    \tau \sim \: \delta k ^{-1/2},
    \label{eq_relaxation_time}
\end{equation}
establishing that the relaxation time diverges near the desynchronization point. The exponent of $-1/2$ is characteristic of a saddle-node bifurcation \cite{kuehn2008scaling}, hinting that formalism described in this Letter can be applied to a broader class of dynamical systems where an abrupt desynchronization occurs along with hysteresis.

Fig.~\ref{figure2} indicates that the deviations from the bulk scaling differ for different $N$  values. Therefore, in addition to slowing down the dynamics, the bulk scaling Eq~\eqref{scaling_r1} requires corrections accounting for the finite size of the system. We calculate a FTFSS of the observable $r(t,N,k)$ using the homogeneity assumption of the classical theory of critical phenomena.

\paragraph{FTFSS:} 
A true critical point exists only in the thermodynamic limit, i.e., when the system size tends to infinity \cite{lee2014finite}. The theory of finite size scaling rests on the assumption that for a finite length $L$, the correlation length $\xi \sim L$. Both $\xi$ and $L$ possess a Lebesgue measure, i.e., the correlations are observed in spatial degrees of freedom. In our mathematical model, the set of all interacting units $I = \{ 1, 2, \dots ,N \}$ posses a counting measure. Consequently the correlation $\xi$ must lie in a counting measurable point set. 
If in the infinite sized system, an exponent $\nu$ defines the divergence in correlation measure $\xi \sim (\delta k)^{-\nu}$, then in finite system $\xi \sim N$. Consequently, $\delta k \sim N^{-1/\nu}$, the relaxation time $\tau \sim N^{\eta/\nu}$ from \eqref{eq_relaxation_time}. This explains the larger deviation for higher $N$ in Fig. \ref{figure2}.
The order parameter $r_s \sim N^{-\beta/\nu}$ where $\beta$  is defined as $\delta r \sim \delta k ^\beta$ in \eqref{scaling_r1}.  
Under the homogeneity assumption, an observable $r(t,N,k)$ is a generalized homogeneous function of all the relevant parameters near the critical point , the scaling ansatz takes the form
\begin{equation}
    r(t,N, \delta k) = N^{-\gamma } \mathcal{R}(N^{-\eta / \nu}t, N^{1/\nu} \delta k).
    \nonumber
\label{ftfss}
\end{equation}
To verify this FTFSS, we perform ensemble averaging since individual realizations from $g(\omega)$ under random sampling fails to exhibit discernible power law behaviour. In contrast, for uniform sampling from $g(\omega)$, we observe a strong system-size-dependent correlation, which renders numerical extraction of scaling exponents computationally prohibitive. A random sampling of frequency is known to cause violation of a hyperscaling relation \cite{hong2015finite}. This caveat will manifest in the numerical results of the amplitude scaling. Nevertheless, we proceed to focus on the theoretical aspects
of the problem which remains robust against this constraint.  We simulate Eq~\eqref{model} for $1000$ realizations and average over a disorder introduced by random sampling of natural frequencies from a Lorentzian distribution.
The temporal and coupling scalings yield exponents $z = \eta/\nu = 0.2$ (Figs.~\ref{figure3}(a) and ~\ref{figure3}(d) ) and $\nu = 2/5$ (Figs.~\ref{figure3}(b) and ~\ref{figure3}(e)), respectively, which agree well with the bulk exponents $\beta = 1/2$ and $\eta = 1/2$. However, the amplitude scaling consistently admits  $\gamma = 0$ value despite that the analytically predicted value  is $\gamma = \beta/\nu =  0.2$. This masking of amplitude exponent may be due to the averaging. Such a mismatch due to averaging has been reported by P\'azm\'andi {\it et. al.} \cite{hong2015finite}. The exponents predicted for other two relevant parameters $t$ and $k$ match well with analytical predictions, and the surfaces for different $N$ collapse  for $\gamma = 0$ near the critical point. The FTFSS performed here suggests the presence of correlation in a number space.

\paragraph{Conclusions and Outlook:}
This study demonstrates that triadic interactions essentially change the scaling characteristics of Kuramoto oscillator systems, adding power-law relations $\delta r \sim \delta k^{1/2}$ and $\tau \sim \delta k^{-1/2}$ that describe the divergences at desynchronization transition. The FTFSS analysis, which is performed in terms of correlations defined via counting measure, determines universal scaling with exponents $z = 0.2$ and $\nu = 2/5$. These findings illustrate that scaling behavior, which is classically linked with equilibrium critical points, can occur in non-equilibrium dynamical systems through the underlying mathematical structure of the governing equations. The observed exponents are consistent with the saddle-node bifurcations that governs the transition dynamics in this system. The scaling relations could be applicable for studying other oscillator systems with first-order explosive transition. Successful application of FTFSS  suggests that these methods may prove useful for characterizing collective behavior in other non-equilibrium systems. This study suggests that presence of non local interactions may allow for critical behaviour at a first order transition. Additional study of these scaling relationships in various network architectures and types of interactions would be informative for discerning their universality and possible uses.

\appendix
\section*{Appendix}
\section{Appendix 1: Bulk scaling through perturbative expansion}
The characteristic function is defined as
\begin{equation}
    F(q) = \frac{q - k_h \: R(q) \: R_2(q)}{R(q)} \equiv \: k\quad,
    \tag{A1.1}
\end{equation}
which has local extrema at all critical points of the system. 

Taking a perturbation $\delta q \ll 1$ in the parameter $q$ at $q_c$ and expanding $F(q_c + \delta q) = F(q_c) + \delta [F(q)]$, we get
\begin{align}
    \delta k &=\: \delta [F(q)] \nonumber \\
             &=\: X \delta q^a  + Y \delta q^b \quad;\: a < b, \tag{A1.2}
\end{align}
where $a$ and $b$ are lowest surviving exponents in the Taylor series expansion of $F(q)$. $a$ and $b$ are hence integers. We can invert this relation to get $\delta k$ dependence on $\delta q$,
\begin{align}
    (\delta k)^{1/a} &= \left[ X \delta q^a  + Y \delta q^b \right]^{1/a}\nonumber \\
    &=X^{1/a} \delta q \left( 1 + \frac{Y \delta q^b}{X \delta q^a}\right) \nonumber \\
    \therefore \quad \delta q \: &=  X^{-1/a} \delta k^{1/a}\left( 1 + \frac{Y \delta q^b}{X \delta q^a}\right)^{-1/a}\nonumber \\
    &= X^{-1/a} \delta k^{1/a}\left( 1 - \frac{Y \delta q^b}{a X \delta q^a}\right) \nonumber \\
    &=X^{-1/a} \delta k^{1/a} - \frac{Y}{a} X^{-\frac{1}{a}-\frac{b}{a} } \:\: \delta k^{\frac{b}{a} + \frac{1}{a} -1} \nonumber \\
    &= X^{-1/a} \delta k^{1/a} + H \: \delta k ^c \tag{A1.3}
\end{align}
with $H = -\frac{Y}{a} X^{-\frac{b}{a} - \frac{1}{a}}$ and $c = \frac{b}{a} + \frac{1}{a} -1$. By comparing the two exponents, it can be seen that $c > \frac{1}{a}$. Therefore, the leading exponent is $1/a$, which, for Cauchy or Gaussian distribution of natural frequency, becomes the mean field exponent $1/2$. 

By definition, $q = k\: r + k_h\: r\: r_2$, which gives a delta variation
\begin{align}
    \delta r &= \frac{1}{k_c + k_h r_{2,c}}\delta q + q_c\: \delta \left( \frac{1}{k_c + k_h r_{2,c}} \right) \nonumber \\
    &= \frac{1}{k_c + k_h r_{2,c}}\delta q + \frac{q_c}{(k_c + k_h r_{2,c})^2}\delta k + \frac{.q_c\: k_h}{k_c + k_h r_{2,c}}\delta r_2. \nonumber
\end{align}
Using the definition of $r_2 = R_2(q)$, we get $\delta r_2 = R_2'(q)\:\delta q$. $R_2(q)$ can be integrated exactly for a density function which is analytic in the unit circle in complex plane \cite{sabhahit2024prolonged},
\begin{align}
    R_2(q) &= \frac{2 + q^2 - \sqrt{q^2 + 1}}{q^2} \nonumber \\
    \therefore \quad \delta r_2 &= \left[ \frac{2 \left( 2 + q^2 -2 \sqrt{q^2 + 1}\right)}{q^3 \sqrt{q^2 +1}} \right] \delta q \nonumber \\
    &= A\:\: \delta q + B \:\: \delta k \nonumber 
\end{align}
with $A = \left[\frac{q_c\;:k_h \: R_2'(q)}{(k_c + k_h r_{2,c})^2} + \frac{1}{k_c + k_h r_{2,c}}  \right]$ and $B = \frac{q_c}{(k_c + k_h r_{2,c})^2}$.

Using the expression for $\delta q$ and $\delta r_2$ in the expression of $\delta r$, we get the bulk amplitude scaling as
\begin{align}
    \delta r = A\: X^{-1/\alpha} \: \delta k ^{1/\alpha} + H\: A\: \delta k ^\epsilon + B\:\delta k. \tag{A1.4}
\end{align}
The exact value of scaling exponents thus depend on the intrinsic frequency distribution; for $g''(w) \neq 0$, $\alpha = 2$ and the model system shows square root scaling which is universal to mean field models. The common frequency distributions are Gaussian and Lorentzian, which fall under this universality class of saddle node bifurcations.
\begin{figure}[H]
    \centering
    \makeatletter
    \renewcommand{\fnum@figure}{}
    \makeatother
    \includegraphics[width=0.5\linewidth]{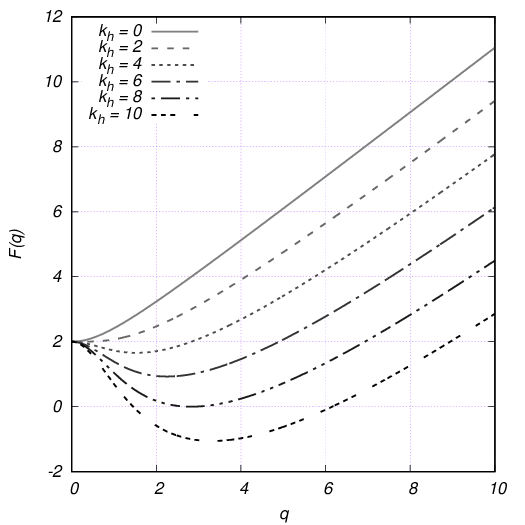}
    \caption{Figure A1.  $F(q)$ is plotted as a function of $q$ for various $k_h$ values.}
    \label{fig:S1}
\end{figure}
Fig.~A1 shows that below a threshold, $q=0$ is a global minimum which is the case for second order transition. For $k_h > k_c$, the minimum of $F(q) $ shifts to non-zero $q$ and induces a first  order transition. In the next section, we will construct a Lyapunov function that provides an energy landscape and illustrates the distinction between the two types of bifurcations.

\section{Appendix 2: Dimensionality Reduction}
Writing model Eq.~(1) as
\begin{equation}
    \label{predict}
	\dot{\theta_i} =\omega_i^{(l)}+  \hat{H}_i
\tag{A2.1}
\end{equation}
where, 
\begin{eqnarray*}
\hat{H}_i =\frac{k}{N}\sum\limits_{j=1}^{N}\sin(\theta_j-\theta_i) 
 + \frac{k_h}{N^2} \sum\limits_{j,l=1}^{N}\sin(2\theta_j-\theta_l-\theta_i)  \nonumber\\
\end{eqnarray*}
 In the mean field, the dynamical evolution equation can be written as
\begin{equation*}
	\dot{\theta_i} =\omega_i+k r\sin(\Psi_1-\theta_i) + k_h r r_2\sin(\Psi_2-\Psi_1-\theta_i) \nonumber \\,
	\label{eq_mean} 
        \tag{A2.2}
	\end{equation*}
with the complex order parameters defined as, 
\begin{align}
z &= r e^{\iota \Psi} = \frac{1}{N}\sum_{j=1}^N e^{\iota \: \theta_j} 
\tag{A2.3}
\\
 z_2&= r_2 e^{\iota\Psi_2}=\frac{1}{N}\sum\limits_{j=1}^{N}e^{2 \:\iota  \: \theta_j}
\label{eq_order}
\tag{A2.4}
\end{align}

Using the complex order parameters \eqref{eq_order}, \eqref{eq_mean} can be written as 
\begin{equation}
 \dot{\theta_i}=\omega_i+\frac{1}{N}(He^{-\iota\theta_i}-H^{*}e^{\iota\theta_i})  
 \label{eq_mean2}
 \tag{A2.5}
\end{equation}
 with $H=k\: z+ k_h\:  z_2\: z^{*}$.
 
 In the thermodynamic limit $N \longrightarrow \infty$, the state of system can be described by a density function $f(\theta,\omega,t)$ which measures the density of oscillators with phase lying between $\theta$ and $\theta + d\theta$ having natural frequency between $\omega$ and $\omega + d\omega$ at time $t$. 
 Since the number of oscillators is conserved, the density function will satisfy the continuity equation,
{\small{
\begin{equation}
    0=\frac{\partial {f}}{\partial t}+\frac{\partial}{\partial \theta} \left[{f} \left[ \omega_i + \frac{1}{N} (H\:e^{-\iota\theta}-H^{*}e^{\iota\theta}) \right] \right]
    \label{eq_continuity}
    \tag{A2.6}
\end{equation}}}
Assuming the natural frequency $\omega$ of each oscillator to be drawn from a distribution $g(\omega)$, the density function $f$ can be expanded into Fourier series as
{\small{
\begin{equation}
    f(\theta,\omega,t) 
    =\frac{g(\omega)}{2\pi} \left[1+\sum\limits_{n=1}^{\infty}\hat{f_n}\:(\omega,t)\:e^{\iota n\theta} + c.c. \right],\nonumber
    \end{equation}}}
where $\hat{f_n}(\omega,t)$ is the $n^{th}$ Fourier component of the density function $f$ and c.c. are the complex conjugates of the former terms. 

\textbf{Theorem:} If a function $f(z)$ is analytic on $\left| z \right| = 1$ and has analytic continuation in an annular region $\mathcal{A} := \{ z \in \mathbb{C} : \: 1-\epsilon  < |z| < 1+\epsilon\}$, the $n^{th}$ frequency mode of Fourier expansion of $f(z)$ decays exponentially with $n$.

One can prove this theorem intuitively as: The basis of Fourier expansion $e^{\iota n \theta }$ diverges exponentially with n and in order to keep the function analytical, the coefficients must counter the divergence. While this intuitive proof is a good starting point, the formal proof reveals the importance of the domain. 

\begin{proof}
Since $f$ is analytic in the annulus $\mathcal{A}$, it admits a Laurent series expansion:
$$f(z) = \sum_{n=-\infty}^{\infty} c_n z^n$$
which converges absolutely and uniformly on any compact subset of $\mathcal{A}$.

The Laurent series can be written as the sum of two parts:
$$f(z) = \sum_{n=0}^{\infty} c_n z^n + \sum_{n=1}^{\infty} c_{-n} z^{-n}$$
where the first sum represents the analytic part (Taylor series) and the second sum represents the principal part. The convergence of this series requires:
\begin{itemize}
\item The Taylor part $\sum_{n=0}^{\infty} c_n z^n$ converges for $|z| < R$ where $R-\epsilon > 1$
\item The principal part $\sum_{n=1}^{\infty} c_{-n} z^{-n}$ converges for $|z| > r$ where $r + \epsilon < 1$
\end{itemize}

The Fourier coefficients are given by the contour integral:
$$c_n = \frac{1}{2\pi i} \oint_{|z|=1} \frac{f(z)}{z^{n+1}} dz$$

Since $f$ is analytic in $A$, we can deform the contour of integration to any circle $|z| = \alpha$ where $r < \alpha < R$. This follows from Cauchy's theorem, as the integrand has no singularities in the region between the two circles.

\textbf{Case 1: $n \geq 0$}

For $n \geq 0$, we choose $\alpha_1$ such that $1 < \alpha_1 < R$. Deforming the contour to $|z| = \alpha_1$:
$$c_n = \frac{1}{2\pi i} \oint_{|z|=\alpha_1} \frac{f(z)}{z^{n+1}} dz$$

Since $f$ is analytic and hence bounded on the compact set $|z| = \alpha_1$, there exists $M_1 > 0$ such that $|f(z)| \leq M_1$ for $|z| = \alpha_1$. Using the standard estimate for contour integrals:
$$|c_n| \leq \frac{1}{2\pi} \cdot 2\pi \alpha_1 \cdot \frac{M_1}{\alpha_1^{n+1}} = \frac{M_1}{\alpha_1^n}$$

Since $\alpha_1 > 1$, we have $\alpha_1^{-1} < 1$, giving exponential decay for $n \geq 0$.

\textbf{Case 2: $n < 0$}

For $n < 0$, let $m = -n > 0$. We choose $\alpha_2$ such that $r < \alpha_2 < 1$. Deforming the contour to $|z| = \alpha_2$:
$$c_{-m} = \frac{1}{2\pi i} \oint_{|z|=\alpha_2} \frac{f(z)}{z^{-m+1}} dz = \frac{1}{2\pi i} \oint_{|z|=\alpha_2} f(z) z^{m-1} dz$$

Again, $f$ is bounded on $|z| = \alpha_2$ by some constant $M_2 > 0$:
$$|c_{-m}| \leq \frac{1}{2\pi} \cdot 2\pi \alpha_2 \cdot M_2 \alpha_2^{m-1} = M_2 \alpha_2^m$$

Since $\alpha_2 < 1$, this gives exponential decay for $m > 0$, i.e., for $n < 0$.

\textbf{Combining the cases:}

Define $\alpha = \max\{\alpha_1^{-1}, \alpha_2\}$. Since $\alpha_1 > 1$ and $\alpha_2 < 1$, we have $\alpha < 1$. Setting $C = \max\{M_1, M_2\}$, we obtain:
$$|c_n| \leq C \alpha^{|n|}$$
for all $n \in \mathbb{Z}$.

This completes the proof of exponential decay.
\end{proof}

Our density function thus follows $\hat{\rho} = \alpha(\omega,t)^n$ with $|\alpha| = |e^{-b}| <1$. This gives
\begin{equation*}
\rho(\omega,\theta,t) = \frac{g(\omega)}{2\pi} \left[ 1 + \left(\sum_{n=1}^\infty \hat{\rho}_n e^{\iota n \theta } + c.c  \right) \right]
\end{equation*}

Putting these Fourier modes into the continuity equation $ \frac{\partial \rho}{\partial t} + \frac{\partial}{\partial \theta}[\;\rho \;\dot{\theta} \;]$, we see that all the modes give the same equation in Fourier space
\begin{equation}
    \dot{\alpha}=-\iota \omega\alpha+\frac{1}{2} \left[H^{*}-H\alpha^{2}  \right]
    \label{eq_alpha}
    \tag{A2.7}
\end{equation} 
with $H$ defined in Eq.~\ref{eq_mean2}. 
The order parameter in the thermodynamic limit can then be given as\\ $z$=$\int \int f(\theta,\omega,t)e^{\iota\theta}d\theta d\omega$, which after inserting the Fourier decomposition of $f(\theta, \omega, t))$ becomes,
\begin{equation}
    z=\int \alpha(\omega,t)g(\omega)d\omega
    \tag{A2.8}
\end{equation}
    If we choose $g(\omega)$ to be a Lorentzian frequency distribution $g(\omega)=\frac{\Delta}{\pi\left[\Delta^{2}+(\omega-\omega_0)^2\right]}$, where $\omega_0$ is mean and $\Delta$ is the standard deviation. $z^{*}$ can be calculated by contour integration in the negative half complex plane, yielding, $z^{*}=\alpha(\omega_0-\iota\Delta,t)$.

\begin{equation}
        \dot{r}=-\Delta r+ \frac{k}{2}r(1-r^2)+\frac{k_h}{2} r^3(1-r^2); \quad \quad
    \dot{\Psi}=\omega_0
    \tag{A2.9}
    \label{final}
\end{equation}
Through dimensionality reduction, the infinite-dimensional system is reduced to this low dimensional form, capturing the essential bifurcation dynamics while maintaining the key physical features. For the numerical analysis and visualization of the bifurcation behavior, we set $\Delta = 1$ in the following plot. This reduced equation provides a comprehensive framework for understanding the global dynamics and transition mechanisms in the thermodynamic system.  In the following figures, the att the equilibrium states of ~\eqref{final} are plotted, stable states in black solid lines and unstable states in red dashed line.
\begin{figure}[H]
    \centering
    \makeatletter
    \renewcommand{\fnum@figure}{}
    \makeatother
    \includegraphics[width=0.35\linewidth]{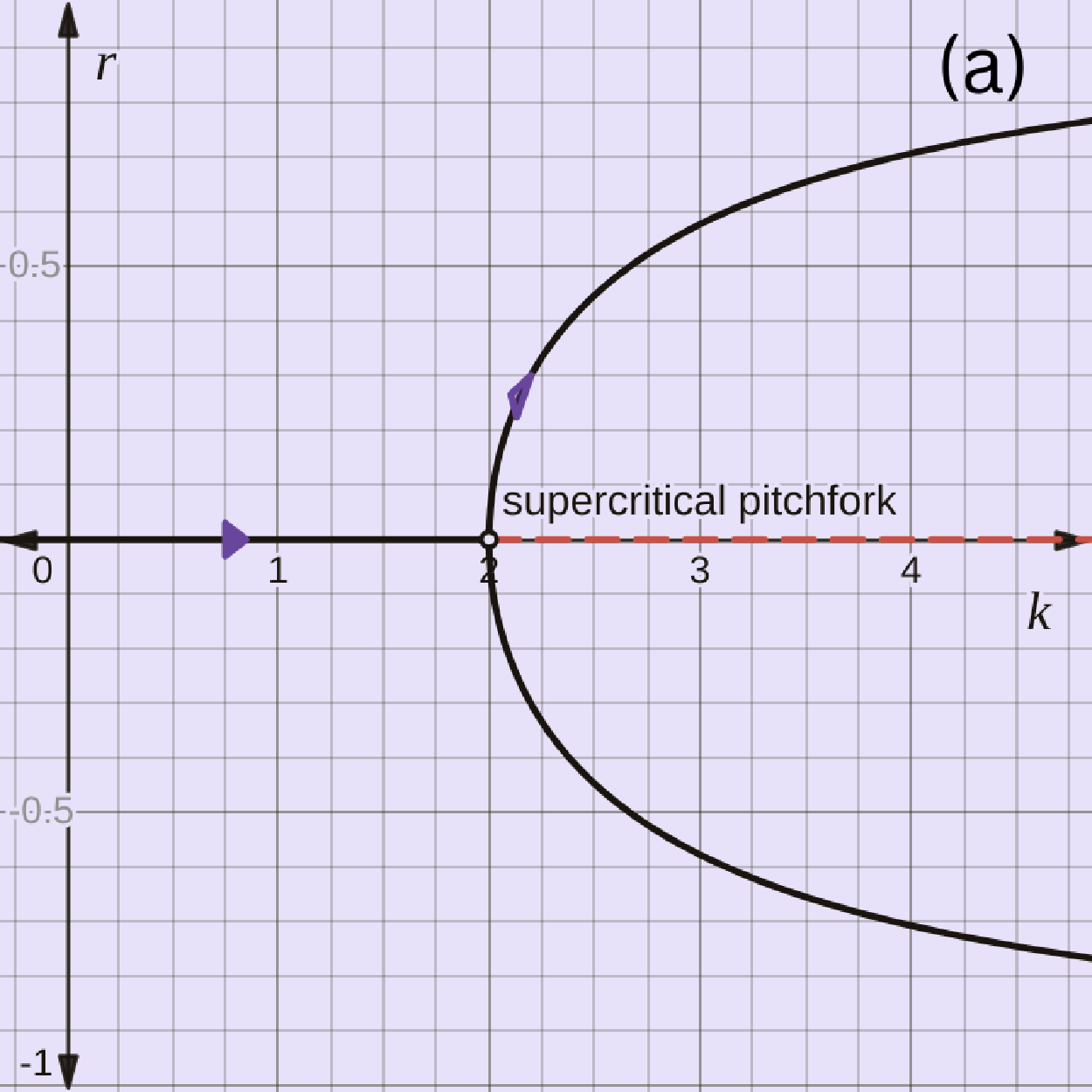}
    \hspace{20pt}
    \includegraphics[width=0.35\linewidth]{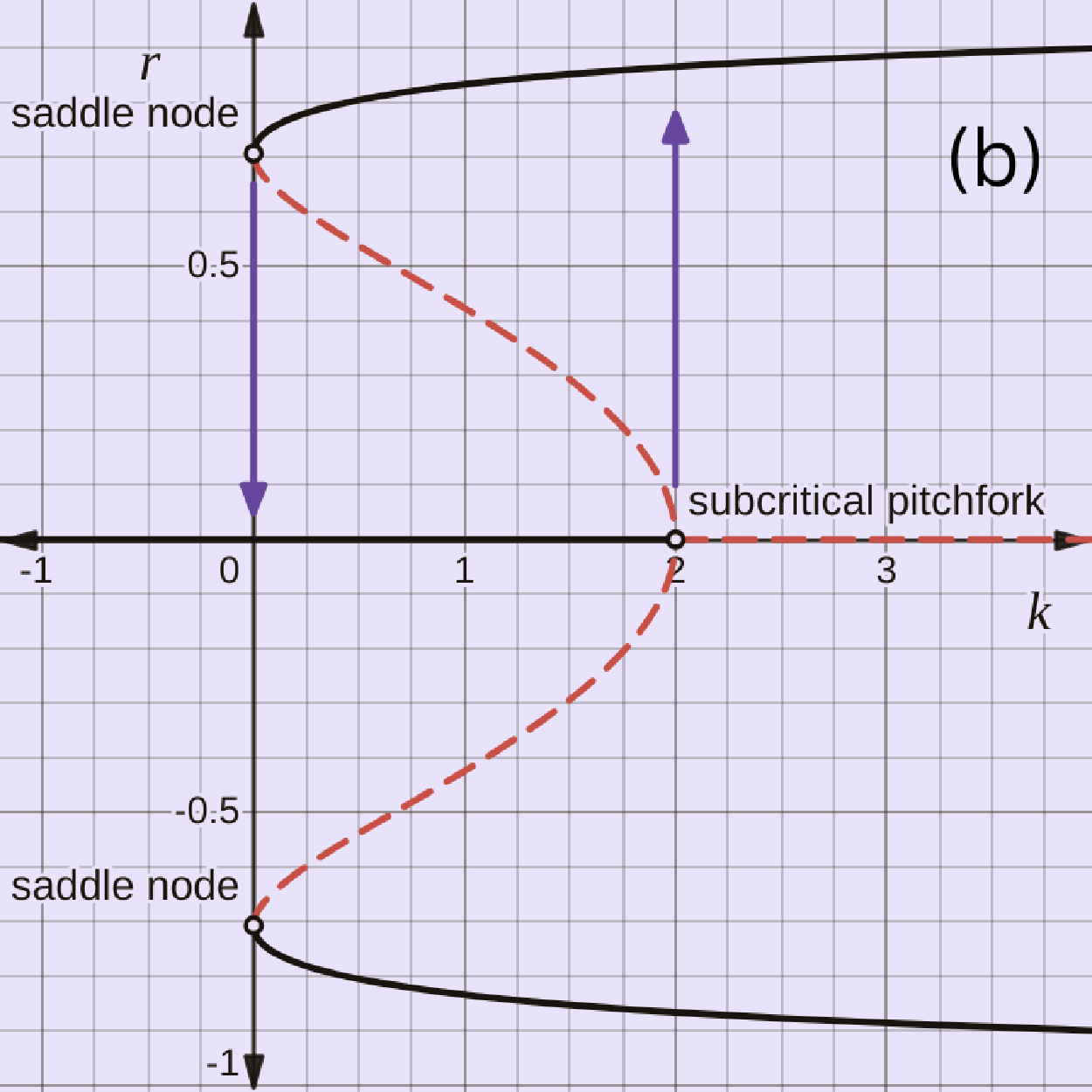}
\caption{Figure A2. Bifurcation diagram. Stable and unstable manifolds for (a) $k_h = 0$, (b) $k_h=8$.}
    \label{fig:enter-label}
\end{figure}
Fig.~A2 reveals the transition between different dynamical regimes as the control parameter varies, considering $r<0$ is in accessible to the system. Panel (a) shows the stable and unstable manifolds for $k_h = 0$, illustrating the supercritical pitchfork bifurcation and a reversible disorder-order transition. Some notable physical examples of this type of transition are: Ising model, Heisenberg model, laser threshold, optical bistability, Eular buckling, and parametric pendulum. Panel (b) demonstrates the corresponding bifurcation structure for $k_h = 8$ (this choice is arbitrary), where the system undergoes a qualitative change in behavior at two critical parameter values. The subcritical bifurcation marks the disorder to order transition via an abrupt jump, while the saddle node bifurcation marks the disorder to order transition via an abrupt jump. It is at this saddle node point, that anomalous behaviour is observed. This bifurcation analysis provides insight into the underlying mechanisms governing the transition between ordered and disordered state through hysteresis. 

SJ and AS acknowledge SERB Power grant (SPF/2021/000136) from Government of India and Council of Scientific and Industrial Research, respectively. We are thankful to Charo Del Genio and Janos Kertesz for useful comments on the work.

\bibliographystyle{apsrev4-2}
\bibliography{refrences} 

\end{document}